\documentclass[prd,onecolumn,notitlepage,nofootinbib,superscriptaddress,showpacs,showkeys]{revtex4-2}

\usepackage{amsfonts}
\usepackage{graphicx}
\usepackage{dcolumn}
\usepackage{amsmath}
\usepackage{amssymb}
\usepackage[T1]{fontenc}
\usepackage[utf8]{inputenc}
\usepackage{esint}
\usepackage[brazil,english]{babel}
\usepackage[usenames,dvipsnames]{color}
\usepackage{longtable}
\usepackage{ulem}
\usepackage{nicefrac}
\usepackage{bm}
\usepackage{tikz}
\usepackage{braket}
\usepackage{graphicx, nicefrac}
\usepackage{hyperref}

\usetikzlibrary[shapes,arrows]

\begin{document}

\title{Black-hole Evaporation for Cosmological Observers}

\author{Thiago de L. Campos}
\email{thiagocampos@alumni.usp.br}
\affiliation{Universidade de S\~{a}o Paulo, Instituto de F\'{\i}sica,
Caixa Postal 66318, 05315-970, S\~{a}o Paulo-SP, Brazil}

\author{C. Molina}
\email{cmolina@usp.br}
\affiliation{Universidade de S\~{a}o Paulo, Escola de Artes, Ci\^{e}ncias e Humanidades, Avenida Arlindo Bettio 1000, CEP 03828-000, S\~{a}o Paulo-SP, Brazil}

\author{J. A. S. Lima}
\email{jas.lima@iag.usp.br}
\affiliation{Universidade de S\~{a}o Paulo, Instituto de Astronomia, Geof\'{\i}sica e Ci\^{e}ncias Atmosf\'{e}ricas, Rua do Mat\~{a}o 1226, CEP 05508-090, S\~{a}o Paulo-SP, Brazil}

\begin{abstract}

This work investigates the evaporation of black holes immersed in a de Sitter environment, using the Vaidya-de Sitter spacetime. The role of cosmological observers is highlighted in the development and Hayward thermodynamics for non-stationary geometries is employed in the description of the compact objects. 
The results of the proposed dynamical model are compared with the usual description based on stationary geometries, with specific results for primordial black holes (PBHs).
The timescale of evaporation is shown to depend significantly on the choice of cosmological observer and can differ substantially from predictions based on stationary models at late times.
Deviations are also shown with respect to the standard assertion that there is a fixed initial mass just below $10^{15} \, \text{g} \sim 10^{-18} M_\odot$ for the PBHs which are completing their evaporation process at the present epoch.

\end{abstract}


\maketitle

\section{Introduction}
\label{sec:level1} 

Black holes are inherently dynamical objects, evolving through either matter accretion or Hawking emission.
While they are often approximated as static in slow-variation regimes, they are never truly stationary.
In addition, the~black-hole background can itself be dynamical~\cite{saida2007black,Faraoni:2008tx,deLima:2010pt,Firouzjaee:2010ia,Moradi:2015caa,Nolan:1998xs,Lake:2011ni,Faraoni:2014nba,carrera,daSilva:2015mja}. For~instance, considering a primordial black hole (PBH) formed during the radiation-dominated era, it will evolve during  the expansion of the Universe, eventually reaching the period when the cosmological constant dominates~\cite{Ruiz:2020yye}. 
This scenario has led to growing interest in PBHs as potential dark matter candidates~\cite{Khlopov:2008qy,carr2016primordial,carr2020primordial,masina2020dark,Green:2020jor,Calza:2024fzo}.

Despite their dynamical evolution, the~evaporation of PBHs is often modeled using the Schwarzschild geometry. These static treatments predict a PBH lifetime on the order of \citep{Belotsky:2014kca,carr2021constraints,calmet2014quantum,Auffinger:2022khh,carr2025history}
\begin{equation} 
	\frac{G^2 M_0^3}{\hbar c^4} \sim 10^{64} \bigg(\frac{M_0}{M_\odot}\bigg)^3 \ \text{yr} \, ,
	\label{typical_time_frame}
\end{equation}
where $M_{0}$ is the initial mass of the PBH. More precisely, Equation~\eqref{typical_time_frame} refers to the proper time measured by a static observer at (asymptotically flat) infinity. This leads to the well-known prediction that PBHs originated with an initial mass smaller than $10^{15}$ g would have already evaporated completely by the present epoch. 
However, in~an accelerating universe, deviations from this timescale are~expected.

In this work, we are interested in the Vaidya-de Sitter spacetime \citep{mallett1985radiating}, which describes a black hole emitting or accreting photons immersed in a universe dominated by a positive cosmological constant. Following the approach used in \citep{hiscock1981models, hiscock1981modelsII}, an~evaporation model for black holes in a de Sitter background is constructed and the evolution of the compact objects is analyzed from the perspective of cosmological observers, which asymptotically comove with the expansion of the spacetime. 
In this framework, Hayward's works \citep{hayward1994general, hayward1996gravitational, hayward1998unified, hayward2009local} provide an important tool as dynamical spacetimes, such as Vaidya-de Sitter, can be thermodynamically treated. 
Underlying this proposal is the fact that the dynamics of a PBH should assume neither a static nor a pure de Sitter description of the Universe. These two scenarios are considered to be the extremes of a spectrum of possible PBH models based on FLRW~cosmology. 

To examine the limitations of the standard evaporation timescale in Equation~\eqref{typical_time_frame}, we propose to explore the other side of the spectrum, where the expansion rate is maximized. The~discrepancies between the models are expected to become more pronounced in the far future, when the de Sitter description should be more appropriate.
Furthermore, cosmological observers hold a privileged status in this framework. The~process of PBH evaporation observed from Earth can be effectively described using the reference frame established by these asymptotically comoving~observers.

Anticipating some key findings of the present work,
our approach is explored and compared with the conventional static framework, challenging previously established results. We demonstrate that the timescales of black-hole evaporation can exhibit significant dependence on the choice of a cosmological observer. Notably, we identify scenarios in which the evaporation is decelerated, with~some black holes never completely evaporating from the perspective of certain cosmological~observers.

\section{Cosmological Observers in Vaidya-De~Sitter}
\label{VdS-geometry}

Vaidya-de Sitter spacetime describes an asymptotically de Sitter and dynamical black hole, taking into account the backreaction of photons being accreted or emitted radially. Its line element, in~outgoing Eddington-Finkelstein coordinates, is given by
\begin{equation}
	ds^2 = - \left[ 1 - \frac{2GM(u)}{c^2r} - \frac{r^2}{a^2} \right] c^2du^2 - 2 c\, du dr + r^2 d\Omega^2 \, , 
	\label{Vaidya-de Sitter metric}
\end{equation}
in which ${a = \sqrt{\nicefrac{3}{\Lambda}}}$ with $\Lambda$ a positive cosmological~constant.

The dynamics of a black hole due to its evaporation can be described by the mass function $M$ of the retarded time coordinate $u$ \citep{hiscock1981modelsII}. On~the other hand, considering cosmological black holes, such as primordial ones, it is important to  be able to measure the evolution of their mass with respect to a cosmic time. 
This requires comparing the Eddington-Finkelstein time coordinate with the cosmological time, which is measured by observers that are comoving with the expansion of the Universe.
However, these observers are typically defined in homogeneous and isotropic spacetimes, and~the presence of a black hole changes this scenario. Our procedure involves first determining a set of cosmological observers in pure de Sitter space, extending their world lines to Vaidya-de~Sitter. 

In pure de Sitter spacetime, cosmological observers are described naturally with the comoving coordinate representation of the line element,
\begin{align}
	\begin{split}
		ds^2 = -c^2 d\tau^2 + a^2 \cosh^2\frac{c \tau}{a} \left( d\chi^2 + \sin^2 \chi d\Omega^2  \right) \, , 
		\label{de Sitter like FLRW}
	\end{split}
\end{align}
which privileges a set of geodesics characterized by $\chi$, $\theta$ and $\phi$ constant, with~$\chi$ ranging from $0$ to $\pi$. These are the world lines of the observers under our consideration, following the spacetime~expansion.

\begin{figure}[h]
%	\centering
	\includegraphics[scale=0.80]{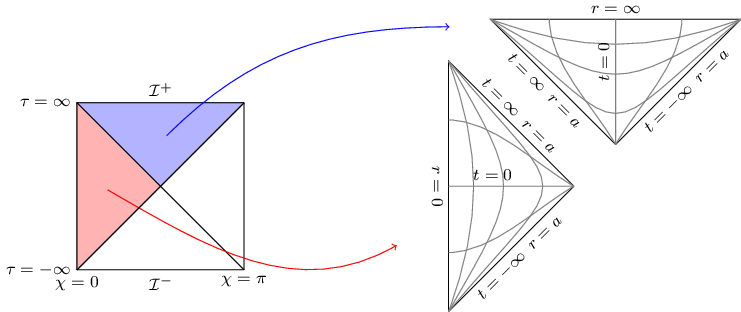}
	\caption{On the left, the~Penrose diagram for the de Sitter spacetime is shown. On~the right, it is highlighted the details of the two disconnected regions that are covered by the $(t, r, \theta, \phi)$ coordinates.}
	\label{fig:Penrose_deSitter}
\end{figure}

In outgoing Eddington-Finkelstein coordinates for de Sitter, the~line element is adapted to the emitted radial null geodesics, assuming the form of Equation~\eqref{Vaidya-de Sitter metric} with a vanishing $M$. We have found the transformation from comoving to outgoing Eddington-Finkelstein coordinates, valid for $0 < \chi < \nicefrac{\pi}{2}\,$, to~be
\begin{equation}
	u' = \, \begin{cases}
		\, \text{arctanh} \bigg( \frac{\tanh \tau' }{\cos \chi} \bigg)
		- \frac{1}{2} \ln \left( \frac{1+r'}{1-r'} \right) 
		+ \frac{1}{2} \ln \left( \frac{1 + \sin \chi}{1- \sin \chi} \right) \,, \quad \textrm{for }0<r'<1 \, ; \\
		
		\,\text{arctanh} \bigg( \frac{\cos \chi }{\tanh \tau' } \bigg)
		- \frac{1}{2} \ln \left( \frac{r'+1}{r'-1} \right) 
		+ \frac{1}{2} \ln \left( \frac{1 + \sin \chi}{1- \sin \chi} \right) \, , \quad \textrm{for }1<r'<\infty \, ;
	\end{cases} 
	\label{transformation cosmic time to u}
\end{equation}
where
\begin{equation}
r = a \cosh \frac{c\tau}{a} \sin \chi \, .
\label{radius from transformation in de sitter}
\end{equation}
In more detail, the~causal structure of the de Sitter spacetime and its two regions that are relevant for the present work are  shown in the Carter-Penrose diagram in Figure~\ref{fig:Penrose_deSitter}.

\begin{figure}[h]
%\centering
\includegraphics[scale=0.8]{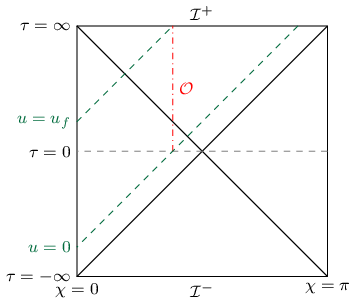}
\caption{Representation of a comoving observer (dash-dotted red  
 line), $\mathcal{O}$, in~the Penrose diagram of the de Sitter geometry. The~dashed green lines indicate the coordinate $u$ at $\tau = 0$ and $\tau \rightarrow \infty$ for the~observer.}
\label{fig:diagram}
\end{figure}

The dimensionless retarded time ($u'$), cosmological time ($\tau'$) and radial position ($r'$) are defined as:
\begin{equation}
u' \equiv \frac{c u}{a} \, , \,\,\,
\tau' \equiv \frac{c \tau}{a} \, , \,\,\,
r' \equiv \frac{r}{a} \, . 
\label{dimensionless radial distance}
\end{equation}
Equation~\eqref{transformation cosmic time to u} was set so that $u' = 0$ when $\tau' = 0$, which explains the restriction on the valid values of $\chi$, as~illustrated by the Carter-Penrose diagram in Figure~\ref{fig:diagram}. As~shown in the diagram, it is clear that $\chi\geq\nicefrac{\pi}{2}$ cannot be set with $u'=0$.

Alternatively, rather than labeling cosmological observers by constant values of $\chi$, it is convenient to label them by $r'_{0}$, where $r'_{0}$ is the value of the dimensionless radius $r'$ when $\tau' = 0\, $. That is, $r'_{0}$ gives the initial position of the observer relative to the cosmological horizon:
\begin{equation}
r'_0 \equiv \sin \chi \, , \hspace{0.5 cm} 0 < r'_0 < 1 \, . \label{spectrum of r0}
\end{equation}

For each value of $r_0'$, the~map
\begin{equation}
		\alpha: \tau \mapsto \left[ u(\tau), r(\tau), \theta_{0}, \phi_{0} \right] 
		\label{cosmological_observers}
	\end{equation}
defines a curve parameterized by the cosmological time in de Sitter spacetime, describing cosmological observers receding exponentially from the coordinate origin. 
At this point, we extend these curves from pure de Sitter to Vaidya-de Sitter. In this setting, the~worldlines of cosmological observers retain the same functional form as in Equation~\eqref{cosmological_observers}, but~are now embedded in the dynamical Vaidya-de Sitter geometry.

\begin{figure}[h]
%		\centering
		\includegraphics[scale=0.8]{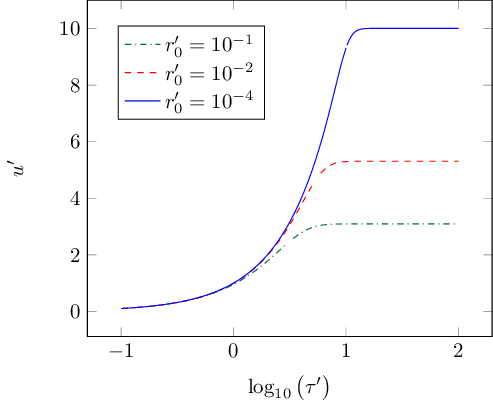}
		\caption{Dimensionless coordinate $u'$ as a function of the dimensionless cosmological time $\tau'$ for different cosmological observers (labeled by the dimensionless initial radial position). From~top to bottom, each curve represents a cosmological observer that reaches the asymptotic regime more quickly according to their proper~time.}
		\label{figure transformation between times}
	\end{figure}

As the observer (labeled by $r'_0$) recedes from the black hole, the coordinate $u$ deviates from the cosmological proper time $\tau$, as illustrated in Figure~\ref{figure transformation between times}.
A direct measurement of the black-hole dynamics with respect to $u$ cannot be correctly compared to the evolution of the background geometry, unless~the transformation between these two coordinates $u$ and $\tau$ is performed.
This point is significant, for instance, when the evaporation of black holes is measured on timescales comparable to the age of the Universe, as it is the case with PBHs.
	
	\section{Evaporation Model}
	\label{dynamic models with VdS}
	
	For the characterization of the evaporation process, the~black-hole temperature must be considered. Given that Vaidya-de Sitter is a dynamical, non-asymptotically flat geometry, Hawking radiation can be suitably framed as a quantum-tunneling phenomenon \citep{hawking1975particle,Birrell,parikh2000hawking}.
	This formalism is carried out in Hayward's generalized thermodynamics \citep{hayward1994general, hayward1996gravitational, hayward1998unified}, where the black-hole boundary is defined as a future outer trapping horizon (FOTH).
	The quasi-local approach near trapping horizons provides a well-defined notion of black-hole temperature in dynamical and multi-horizon settings. In~this formalism, the~temperatures associated with the black hole and cosmological horizon are derived independently. This ensures that the evaporation timescale is unaffected by the cosmological horizon in treatments where thermal exchange between both horizons is~negligible.
	
	Within this framework, the~temperature $T$ associated to the black hole is proportional to the geometric surface gravity $\kappa_G$ of the FOTH \citep{hayward2009local},% 
\begin{equation}
		T(u) = \frac{\hbar\, c}{k} \, \frac{\kappa_G(u)}{2\pi} 
		\, .
		\label{temperature}
	\end{equation}
	For Vaidya-de Sitter geometry, the~geometric surface gravity is given by
\begin{equation}
		\kappa_G(u) = \frac{1}{2} \partial_r \bigg[ 1 - \frac{2GM(u)}{c^2r} - \frac{r^2}{a^2} \bigg]_\text{\tiny FOTH} \, .
		\label{eq:kappa}
	\end{equation}
	The mass loss of the black hole, due to Hawking radiation, is a quasi-local process occurring near its horizon. As~a result, we expect the Schwarzschild term $\nicefrac{2GM}{c^2 r}$ in the metric to dominate the geometry in this region. In~fact, for~reasonable values of the black-hole mass (even for supermassive ones, up~to the order of $10^{11} M_\odot$), the~last term can be neglected at first order due to the extremely small value of the cosmological constant. A~similar analysis leads to the position of the FOTH being well approximated by the (dynamical) Schwarzschild radius, $\nicefrac{2 G M(u)}{c^2}$. This approximation reduces Equation~\eqref{temperature} to the standard expression for the Vaidya black hole \citep{piesnack2022vaidya}.
	
	Combining results~\eqref{temperature} and \eqref{eq:kappa} with the Stefan-Boltzmann law
	\citep{piesnack2022vaidya, carballo2018viability}, the~mass variation of the Vaidya-de Sitter black hole is well approximated by
\begin{equation}
		\frac{dM(u)}{du} = - \frac{\hbar c^4}{15360 \pi G^2} \frac{1}{M(u)^2} \, . 
		\label{mass dynamic evaporation}
	\end{equation}
	Solving this differential equation and applying the coordinate transformation~\eqref{transformation cosmic time to u}, the~mass function $M$ in terms of the dimensionless cosmological time $\tau'$ is determined,

\begin{align}
		\begin{split}
			M(\tau',r'_0,M_0) = \begin{cases}
				\left\{ M_0^3 - \frac{\hbar a c^3}{5120 \pi G^2} \left[ \text{arctanh} \bigg( \frac{\tanh \tau' }{\cos \chi} \bigg)
				- \frac{1}{2} \ln \left( \frac{1+r'}{1-r'} \right) 
				+ \frac{1}{2} \ln \left( \frac{1 + \sin \chi}{1- \sin \chi} \right) \right] \right\}^{\frac{1}{3}} \,, 
				\quad \textrm{for }0<r'<1 \, ;\\   
				\left\{ M_0^3 - \frac{\hbar a c^3}{5120 \pi G^2}  \left[ \text{arctanh} \bigg( \frac{\cos \chi}{\tanh \tau' } \bigg) 
				- \frac{1}{2} \ln \left( \frac{r'+1}{r'-1} \right) 
				+ \frac{1}{2} \ln \left( \frac{1 + \sin \chi}{1- \sin \chi} \right) \right] \right\}^{\frac{1}{3}} \, , 
				\quad \textrm{for }1<r'<\infty \, .
			\end{cases} 
			\label{M(tau)}
		\end{split}
	\end{align}
A characteristic mass for the evaporation model is given by
\begin{equation}
		\frac{\hbar a c^3}{5120 \pi G^2} \simeq \left(9.560 \times 10^{-20} M_{\odot} \right)^{3}~, \label{characteristic mass}
		\end{equation}
		 which can be used as a numerical reference. The~solar mass is, approximately, \mbox{$1.988 \times 10^{30}$ {kg}} and the cosmological constant $\Lambda$ is taken to be $ 10^{-52}$ {m}$^{-2}$ \cite{navas2024review}.
		
		Actually, black-hole evaporation is not perfectly described by a black-body spectrum. Semiclassical treatments reveal that it depends on the species of particles emitted, which in turn depends on the size of the compact object~\cite{calmet2014quantum,carr2021constraints}. Nonetheless, the~Stefan-Boltzmann law is a commonly used approximation, which is particularly useful for the present work as our main results emerge from measuring the evaporation with the cosmological time. As~it will be presented in Section~\ref{analysis}, our model suppresses these quantum corrections to Equation~\eqref{mass dynamic evaporation}, which have no similar~effect.
		
	From the evaporation model proposed in this section, the~choice of a de Sitter background is justified. 
    The evaporation process can be more easily treated when compared to more general cosmological black-hole solutions (e.g., McVittie-type metrics~\cite{Lake:2011ni,Faraoni:2014nba,carrera,daSilva:2015mja,Ruiz:2020yye}). Specifically,  the~Vaidya solution~\eqref{Vaidya-de Sitter metric} explicitly describes a spherically symmetric geometry with a radial null-energy flux $T^r{}_u$. Thus, the~Einstein equations yield a simple relation between the energy-momentum tensor and the rate of change in the Misner-Sharp mass $M_\text{MS}$,
\begin{equation}
	4\pi r^2 T^r{}_u = \frac{dM_\text{MS}}{du} = \frac{d}{du} \bigg( \frac{GM}{c^2} + \frac{r^3}{2a^2} \bigg)  \, .
	\end{equation}
	
	The first equality is valid for any spherically symmetric metric in Eddington-Finkelstein-like coordinates~\cite{campos2025dynamical}, and~the second one is valid for the Vaidya-de Sitter spacetime.
    At the FOTH, the~Misner-Sharp mass provides a precise representation of the black-hole mass, and~the component $T^r{}_u$ thus describes the black hole's emission as it evaporates~\cite{israel1986formationy}. 
	The Vaidya-de Sitter spacetime, which is naturally expressed in Eddington-Finkelstein-like coordinates,
    allows one to directly model the backreaction of Hawking radiation via a Stefan-Boltzmann-type law, relating the black hole temperature to $T^r{}_u$ \cite{balbinot1989backreaction}.

It should be emphasized that the main distinction between the evaporation rate in expanding and stationary universes does not originate from the cosmological constant's direct effect on black-hole temperature. Rather, it stems from its impact on observer trajectories and on the asymptotic structure of spacetime, specifically from the fact that conformal infinity is~spacelike. 

In Schwarzschild spacetime, a distant static observer's proper time approximately coincides with the coordinate $u$, leading to the standard description of evaporation via $M(u)$ and resulting in Equation~\eqref{typical_time_frame}. However, in~a rapidly expanding universe, cosmological observers are more naturally described by the time $\tau$. We therefore propose that Equation~\eqref{M(tau)} provides a more physically appropriate description in this~context.
		
Moreover, for curves close to the horizon%
\footnote{Nonetheless, these curves may not correspond to appropriate world lines for cosmological observers, in~general.},
the coordinate transformation of Equation~\eqref{transformation cosmic time to u} is trivial,~$u \simeq \tau$. For~Equation~\eqref{M(tau)}, using $\tau = \nicefrac{a \tau'}{c}$ and keeping only the function dependence on the cosmological time explicit for simplicity, 
\begin{equation}
M(u) = \left(  M_0^3 - \frac{\hbar c^4}{5120 \pi G^2} u \right)^{\frac{1}{3}} 
\simeq \left(  M_0^3 - \frac{\hbar c^4}{5120 \pi G^2} \tau  \right)^{\frac{1}{3}} \simeq M(\tau)  \, ,  \hspace{0.3cm}\textrm{for } r' \simeq 0 \, . \label{M(tau) tends to usual}
\end{equation}
Equation~\eqref{M(tau) tends to usual} corresponds to the standard model for Hawking evaporation. Therefore, as~one might intuitively expect, our model tends towards the usual treatment near the black hole, where the compact object dominates the~geometry.

		\section{Analyzing the Evaporation Model}
		\label{analysis}
		
		In order to characterize the evaporation rate of the black hole, one must specify both the cosmological observer's initial position $r'_0$ and the initial black hole mass $M_0$. We have conducted a systematic exploration of this parameter space and present here the most representative results. Our analysis focuses on highlighting the main differences between our dynamical model and the conventional static description commonly used for PBHs \citep{carr2016primordial, carr2020primordial, masina2020dark, carr2021constraints, Auffinger:2022khh}.
		When considering scenarios where the dynamical nature of the Universe cannot be overlooked, our results lead to non-standard evaporation~timescales.

The first step is to analyze the accuracy with which the observers defined by Equation~\eqref{cosmological_observers} are actually ``cosmological'', in~the sense that their proper time is approximately equal to $\tau$. Since the black hole is evaporating and the observers are receding, it is sufficient to evaluate the norm of $\dot \alpha$ at the initial time. Ultimately, if~the approximation is accurate enough at $\tau = 0$, it will be even better for latter times. For~the largest black hole investigated in this section ($M_0 = M_\odot$) and choosing an observer with world line relatively close to the black hole,
\begin{equation}
r'_0 = 10^{-10} \implies c^{-2}|\dot \alpha^\mu \dot \alpha_\mu| \geq 1 - 10^{-13}~.
\end{equation}

This reveals a remarkable agreement between this observer's proper time and $\tau$. 
Selecting the correct observer depends on the desired level of precision. Although~exact agreement between proper time and $\tau$ is only achieved asymptotically as $r \to \infty$, our results show that the range of suitable observers is broader than one might assume. 
This implies that one can describe the dynamics of a primordial black hole without fine-tuning the observer and using a proper time that is meaningful in a cosmological~context.

\begin{figure}[t]
%			\centering
			\includegraphics[scale=0.8]{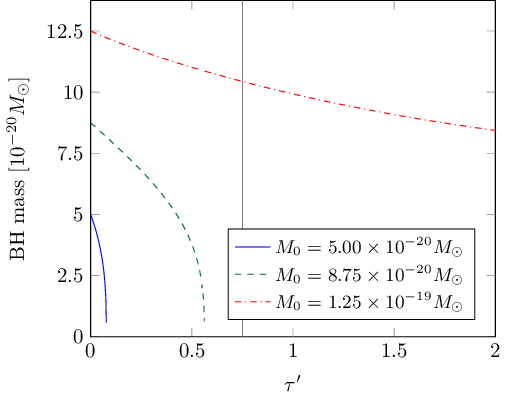}
			\caption{Black-hole mass as a function of dimensionless cosmological time for different values of initial masses. An~observer characterized by ${r'_0 = 0.866}$ is considered. The~vertical line indicates the present epoch (${\tau' \sim 0.75}$, while $\tau'$ is varied.) The curve where ${M_{0} = 1.25 \times 10^{-19} M_{\odot}}$ exemplifies a scenario in which a PBH would still be evaporating today. The other two curves exhibit behaviors similar to those of the traditional Schwarzschild-based~model.}
			\label{m(tau) chi=pi/3}
		\end{figure}

We consider now the evolution of the black hole according to the cosmological observers.  
Figure~\ref{m(tau) chi=pi/3} shows the evolution of $M$ according to our proposed evaporation model, where the curves were derived from Equation~\eqref{M(tau)} with $M_0$ and $r'_0$ fixed. 
The two lighter black holes in this figure (with initial masses of $5.00 \times 10^{-20} M_{\odot}$ and \mbox{$M_0 = 8.75 \times 10^{-20} M_{\odot}$}) exhibit a similar evaporation rate (both numerically and in functional form) to that of the traditional static model, which relies on the Schwarzschild time and follows ${(M^3 - M_0^3) \propto t}$. In~particular, they evaporate completely ($M=0$) at finite values of $\tau'$. On~the other hand, the~heavier black hole ($M_0 = 1.25 \times 10^{-19} M_{\odot}$) exhibits a completely new behavior in terms of its evaporation rate [note that this threshold is of the same order as the characteristic mass~\eqref{characteristic mass}]. Thus, a~key finding is that black holes in static and dynamical backgrounds can evaporate with significantly different profiles, depending on their initial mass and the considered observer.

In fact, from~the perspective of a cosmological observer (for instance, an~observer on Earth), mass loss can occur at a decelerated rate, rather than at the standard accelerated rate of evaporation.
This dynamical effect predicted by the model is illustrated in Figure~\ref{m(tau) chi=pi/3}, as~the curve of positive concavity.
At the present epoch ($\tau' = 0.75$), its mass has only decreased to approximately $1.0 \times 10^{-19} M_{\odot}$. A~standard calculation in the Schwarzschild-based model would show a much larger fractional mass loss over the same time interval.
Remarkably, a~black hole can retain a finite residual mass indefinitely in this~scenario.

When referring to observers on Earth, we assume that PBHs are observed in the early Universe. In~such cases, due to the large cosmological distances involved, the~effects of accelerated expansion play a more significant role in the observational predictions.
		As shown by the dash-dotted red line in Figure~\ref{m(tau) chi=pi/3}, a~black hole with initial mass \mbox{$M_0 = 1.25 \times 10^{-19} M_\odot$} observed by $r'_0 = 0.866$ ($\chi = \pi/3$) approaches the asymptotic mass
\begin{equation}
			\begin{split}
				\lim_{\tau' \rightarrow \infty} M(\tau', r'_0 = 0.866&, M_0 = 1.25 \times 10^{-19} M_\odot) \simeq 7 \times 10^{-20} M_\odot 
				\label{ex1}
			\end{split}
		\end{equation}
		as $\tau' \rightarrow \infty$, never fully~evaporating.
		This effect becomes even more pronounced for more massive black holes. When $M_0 \sim M_\odot$, cosmological observers given by ${r'_0 = 0.866}$ detect essentially no evaporation:
\begin{equation}
			\lim_{\tau' \rightarrow \infty} M(\tau') \simeq M_\odot \, .
			\label{ex2}
		\end{equation}
		The radiation rate becomes so slow from this particular cosmological perspective that the black hole appears effectively~stable.
		
		These results can be understood by examining the relation between cosmological time and the coordinate $u'$.
		The cosmological time extends indefinitely for a finite interval of $u'$. More precisely,
\begin{equation}
			\lim_{\tau' \rightarrow \infty} u' = \text{arctanh} \left( \cos \chi \right) + \frac{1}{2} \ln \left( \frac{1 + \sin \chi}{1 - \sin \chi} \right) = \textrm{finite}\, ,
			\label{lim u = finite value}
		\end{equation}
		reflecting the spacelike character of future infinity in de Sitter spacetime.
		This asymptotic behavior is also apparent in Figure~\ref{figure transformation between times}, where we observe the characteristic ``stretching'' of cosmological time relative to the coordinate $u$.
		
		Our results also indicate that the evaporation rate of a black hole, measured with respect to a cosmological time, is sensitive to its initial mass. Even within the same order of magnitude, for~example at $\text{M}_0 \sim 10^{-19} M_\odot$, we see that completely different behaviors emerge, as~shown in Figure~\ref{m(tau) chi=pi/3}.
		This phenomenon has no counterpart in the usual analysis considering the Schwarzschild~spacetime.
		
		In our framework, the~complete evaporation of black holes becomes an event inaccessible for some observers.
		As the primary objective of this work is to compare our results with Equation~\eqref{typical_time_frame} from traditional static models, we do not take into account strong quantum corrections in the final stages of evaporation, which could potentially lead to relics.
		For the family of cosmological observers that we consider, the~asymptotic black-hole mass can be determined by evaluating the limit:
\begin{equation}
			\lim_{\tau'\rightarrow \infty} M(\tau')  \,\,  \text{for some specific observer $r'_0$}\, .
			\label{eq:observer-dependence}
		\end{equation}
		Prescription~\eqref{eq:observer-dependence} determines the asymptotic black-hole mass at future infinity as a function of both its initial mass $M_0$ and observer's initial position $r'_0$.

\begin{figure}[t]
%			\centering
			\includegraphics[scale=0.8]{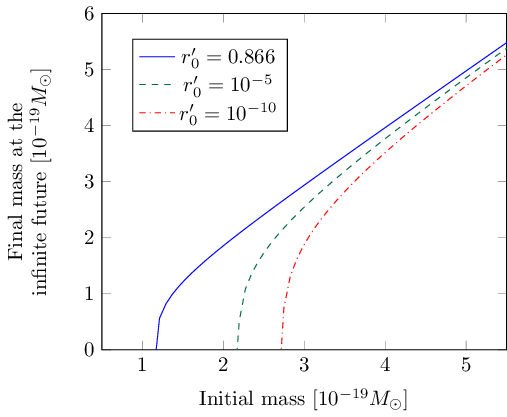}
			\caption{Asymptotic mass of an evaporating black hole at the infinite future for distinct cosmological observers (labeled by their initial radial positions). From~left to right, each curve corresponds to a cosmological observer for which an asymptotically vanishing black hole has a larger initial~mass.}
			\label{final x initial mass}
		\end{figure}

This issue is analyzed in Figure~\ref{final x initial mass}, which quantifies the observer-dependence of the final state of an evaporating black hole. Each curve in the figure follows from the analytical result of Equation~\eqref{M(tau)} with $\tau'\rightarrow \infty$ and $r'_0$ fixed, while $M_0$ is varied. 
For initial masses around $10^{-19} M_\odot$, the~asymptotic final mass is particularly sensitive to the choice of cosmological observer.
For more massive black holes ($M_0 \gtrsim 5 \times 10^{-19} M_\odot$), evaporation rates become largely independent of the observer. This occurs because black holes evaporate extremely slowly, except~for those with very small masses. As~shown in Figure~\ref{figure transformation between times}, in~this case cosmological observers have sufficient time to reach the asymptotic regime, in~which the evaporation rate is significantly reduced from their~perspective.
		
\begin{figure}[h]
%			\centering
			\includegraphics[scale=0.8]{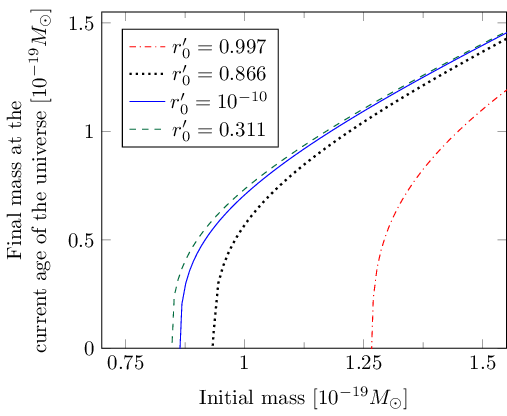}
			\caption{Final mass of an evaporating black hole at the current age of the Universe for distinct cosmological observers (labeled by their dimensionless initial radial position). From~left to right, each curve corresponds to a cosmological observer for whom a black hole that is currently reaching the end of its evaporation process had a larger initial~mass.}
			\label{final x initial mass at current age}
		\end{figure}

		One objective of our analysis is to identify which black holes have completed their evaporation during the current cosmic epoch as observed by our family of cosmological observers. Specifically, we evaluate:	
\begin{equation}
			M(\tau' \simeq 0.75)  \,\,  \text{for some specific observer $r'_0$ }\, ,
		\end{equation}
		where $\tau' \simeq 0.75$ corresponds to the current age of the Universe ($\tau \simeq 4.35 \times 10^{17}$ s) in dimensionless time units.
		This analysis is presented in Figure~\ref{final x initial mass at current age}, where the analytical result~\eqref{M(tau)} is considered with $\tau' = 0.75$ and $r'_0$ fixed, while $M_0$ is varied.
		Several features emerge from the results in this~figure.       
		\unskip

		First, the~evaporation dynamics governed by Equation~\eqref{M(tau)} tends to the standard Schwarzschild description as $r'_0 \to 0$. 
		For example, a~trajectory with $r'_0 = 10^{-10}$ at present cosmic time ($\tau' = 0.75$) matches the Schwarzschild solution with a discrepancy of approximately $10^{-10}$ [verifiable via Equation~\eqref{transformation cosmic time to u}, and~noting that we are comparing analytic solutions].
		Consequently, the~solid blue curve in Figure~\ref{final x initial mass at current age} serves as an effective reference for comparing all other cases with the conventional Schwarzschild evaporation~scenario.
		
		Additionally, the~evaporation rate cannot be simply inferred by the initial distance between the black hole and the observer. For~example, the~evaporation may occur at a slower rate for some observers, as~it happens for $r'_0 = 0.311\,$. In~this case, a~black hole that should have vanished (when considering the usual treatment) could still be present.
		Conversely, observers who start their trajectory close to the cosmological horizon 
		($r'_0 \approx 1$) can experience dramatically accelerated evaporation.
		A striking example occurs for $r'_0 = 0.997$. While a black hole with $M_0 = 1.25 \times 10^{-19} M_\odot$ (red dash-dotted curve in Figure~\ref{final x initial mass at current age}) would have completely evaporated from this observer's perspective, it would show a mass loss of less than 20\% in the Schwarzschild approximation (solid blue curve in Figure~\ref{final x initial mass at current age}).
		        
		As a direct consequence of our analysis, if~a PBH signal were detected, inferring its initial mass from the standard evaporation model could be significantly erroneous, as~our relative motion to the compact object would not be taken into account. For~example, a~detected PBH with a current mass of $5 \times 10^{-20} M_\odot$ could be interpreted as a heavier object in our model, when compared to the standard (static) scenario summarized by Equation~\eqref{typical_time_frame}. 		
		
		\section{Final Remarks}
		\label{remarks}
		
		In this work, we analyze the evaporation of black holes in a de Sitter environment, comparing our results with those from the usual static frameworks, with~special attention to PBHs. 
		In this description, the~Vaidya-de Sitter spacetime is used to model a black hole in an accelerated Universe, explicitly accounting for the crucial role of cosmological observers.
		Building on Hayward's thermodynamic formalism for dynamical spacetimes, our analysis reveals striking departures from conventional evaporation rates. Notably, cosmological observers measure significantly modified PBH evaporation rates compared to static spacetime~predictions.
		
		In several studies examining the evaporation of PBHs, the~phenomenon is investigated without considering the effects of a positive cosmological constant. In~such cases, the~Schwarzschild solution is typically employed as the framework for describing the evaporation process.
		However, if~the spacetime is dynamical and not asymptotically flat, the~preferred observers must be redefined.
		We focus on a class of observers that are (asymptotically) comoving with the expansion of the Universe. For~these observers, we have shown that the rate of evaporation of a black hole can be highly distinct when compared to the Schwarzschild~counterpart. 
		
		In the conventional descriptions, primordial black holes that reach the end of their lifetime at the current age of the Universe are those with an initial mass just below $10^{15} \, \text{g} \sim 10^{-18} M_\odot$.
		Our analysis predicts deviations from this result for a wide set of observers who measure the cosmological time, leading to primordial black holes evaporating earlier or later than expected.
		Naturally, these results must be taken with a grain of salt since a more accurate setting for the dynamics of a PBH should be neither static nor de Sitter but~somewhere in between. This notwithstanding, our model provides a preliminary estimate of the deviation that can be expected when the spacetime expansion is taken into~consideration.
		
		As the Universe ages and its expansion rate accelerates, differences between the commonly used static evaporation model and our dynamic description become more pronounced. 
		As the cosmological constant becomes dominant, the~expected phenomenology asymptotically approaches the description presented in this work.
		One remarkable new result is the fact that, for~certain cosmological observers, complete evaporation is inaccessible and therefore never observed. This is a consequence of the spacelike nature of the future conformal infinity and~represents a significant departure from conventional evaporation scenarios.
		The results of this work challenge conventional expectations based on static models. 
		Nevertheless, they are a genuine consequence of the difference between analyzing evaporation using the coordinate $u$ versus the cosmological time, measured by asymptotically comoving~observers.
		
		This work suggests several new research directions. For~instance, one could extend the analysis to realistic cosmological backgrounds that transition between radiation, matter, and~dark energy cosmological eras, possibly using a McVittie-type solution. Furthermore, investigating the implications for cosmological phenomenology, particularly primordial black-hole constraints, would be valuable. These possibilities are currently under investigation.

\section*{Funding}
    
	T.~L.~C. acknowledges the support of Coordena\c{c}\~ao de Aperfei\c{c}oamento de Pessoal de N\'{\i}vel Superior (CAPES) -- Brazil, Finance Code 001.
	C.~M. is supported by Grant No.~2022/07534-0, S\~ao Paulo Research Foundation (FAPESP), Brazil.
	J.~A.~S.~L. is partially supported by National Council for Scientific and Technological Development (CNPq), Brazil, under Grant 310038/2019-7, CAPES (88881.068485/2014), and FAPESP (LLAMA Project No.~11/51676-9).

\section*{Data Availability Statement}

No new data were created in the course of this work.

\end{document}